\documentclass[english,conference]{IEEEtran}
\usepackage[T1]{fontenc}
\usepackage[utf8]{luainputenc}
\usepackage{amsmath, psfrag} 
\usepackage{amssymb}  
\usepackage{textcomp,relsize}
\usepackage{graphicx}
\usepackage[export]{adjustbox}
\usepackage{subcaption}
\usepackage{balance}
\makeatletter

\providecommand{\tabularnewline}{\\}
\usepackage{textcase}
\usepackage[tablename=TABLE]{caption}
\DeclareCaptionTextFormat{up}{\MakeTextUppercase{#1}}
\usepackage{babel}
\makeatother
\begin{document}

\title{Congestion costs incurred on Indian Roads:\\ A case study for New Delhi}

\author{Neema Davis, Harry Raymond Joseph, Gaurav Raina, Krishna Jagannathan\\
 Department of Electrical Engineering, Indian Institute of Technology
Madras, Chennai 600 036 \\
 E-mail: \{ee14d212, ee10b127, gaurav, krishnaj\}@ee.iitm.ac.in}
\maketitle
\begin{abstract}
We conduct a preliminary investigation into the
levels of congestion in New Delhi, motivated by concerns due to rapidly growing vehicular congestion
in Indian cities. 
First, we provide statistical evidence for the rising congestion levels on the roads of New Delhi from taxi GPS traces. Then, we estimate the economic costs of congestion in New Delhi. 
In particular, we estimate the marginal and the total costs of congestion. 
In calculating the marginal costs, we consider the following factors: (i) productivity loss, (ii) air pollution costs, and (iii) costs due to accidents.
In calculating the total costs, in addition to the above factors,
we also estimate the costs due to the wastage of fuel. 
We also project the associated costs due to productivity loss and air pollution till 2030. The projected traffic congestion costs for New Delhi comes around 14658 million US\$/yr for the year 2030. 
The key takeaway from our current study is that costs due to
productivity loss, particularly from buses, dominates the overall
economic costs. Additionally, the expected increase in fuel wastage
makes a strong case for intelligent traffic management systems. 
\end{abstract}
\begin{IEEEkeywords}
road congestion, marginal cost, total cost, projections 
\end{IEEEkeywords}

\section{Introduction}


Traffic congestion in Indian cities is visibly on the rise. This has
a detrimental effect on productivity, air pollution, fuel wastage,
health, and quality of life. In the developed world, traffic congestion
has long been recognized as an economic as well as a social impediment,
and detailed studies on the economic aspects of congestion have been
conducted. Such studies have been successful in sparking numerous
policy deliberations, and have generated interest in devising novel
traffic management systems.

A brief overview of road congestion statistics in some developed economies
is given below. 
\begin{itemize}
\item Annual congestion cost in the United Kingdom (UK) will reach 33.4
billion US\$ by $2030$, rising by over 50\% from the 2014 levels
of 20.5 billion US\$ \cite{carey}. 
\item Annual cost of congestion in the United States (US) as of 2014, has
been pegged at 124 billion US\$; this is projected to increase to
186 billion US\$ by 2030 \cite{carey}. 
\item In Australia, annual congestion cost levels are expected to rise from
Australian Dollars (AUD) 3.5 billion (2005) to AUD 7.8 billion (2020)
for Sydney, and AUD 3.0 billion (2005) to AUD 6.1 billion (2020) for
Melbourne \cite{police}. 
\end{itemize}
Such extensive studies have not been conducted for Indian cities as
yet. However, it is being recognised that as India develops, congestion
in cities is going to increase sharply, with numerous negative implications.
The following statistics provide some insights into the congestion
scenario in New Delhi: 
\begin{itemize}
\item New Delhi's vehicular population is projected to rise to 10 million by
2020, leading to a marked increase in congestion, which will severely
impede economic activity \cite{police}. 
\item In New Delhi, at least about 300,000 US\$ worth of fuel was being wasted
everyday, by vehicles idling at traffic signals as early as in 1998
\cite{plan}. This figure jumped to approximately 1.6 million US\$ per
day as of 2010 \cite{muller}. 
\item New Delhi has been named the world's most polluted city among 1600 cities
by the World Health Organisation (WHO), and vehicular emissions are
a major contributor to this situation \cite{sarma}.


\end{itemize}

Most of the research in the literature studying economic aspects of
congestion, uses the link-flow approach. An example of this is \cite{muller},
which uses analytical models to establish congestion costs against
a baseline scenario. Another related paper \cite{gris} uses a similar
approach to estimate total traffic congestion costs. The approach
followed by most of these researchers is to use an exponential congestion
function, which relates the minutes needed to drive a kilometer in
terms of the Passenger Car Units (PCU) in the city. We note that previous
studies have not explicitly considered the effect of two-wheelers
on the congestion costs. This may undermine the congestion estimates,
as two-wheelers already outnumber cars, and will also increase in
the future. The impact of two-wheelers has been incorporated in our
study.

\begin{figure*}[h!]
                \begin{subfigure}{0.33\textwidth}
                    \psfrag{d}{\hspace{0mm}\raisebox{-2mm}{\footnotesize{Day}}}
       				\psfrag{s}{\hspace{-3mm}\raisebox{6mm}{\footnotesize{km/h}}}
      				\psfrag{j13}{\hspace{0mm} \raisebox{0mm}{\scriptsize{Jan 2013}}}
      				\psfrag{j14}{\hspace{0mm} \raisebox{-0.5mm}{\scriptsize{Jan 2014}}}
      				\psfrag{0}{\hspace{0mm} \raisebox{-2mm}{\footnotesize{0}}}
      				\psfrag{31}{\hspace{-3mm} \raisebox{-2mm}{\footnotesize{31}}}
      				\psfrag{30}{\hspace{-5mm} {\footnotesize{30}}}
      				\psfrag{32.5}{\hspace{-6mm} {\footnotesize{32.5}}}
      				\psfrag{35}{\hspace{-5mm} \raisebox{-2mm}{\footnotesize{35}}}
      
     				 \includegraphics[width=\linewidth]{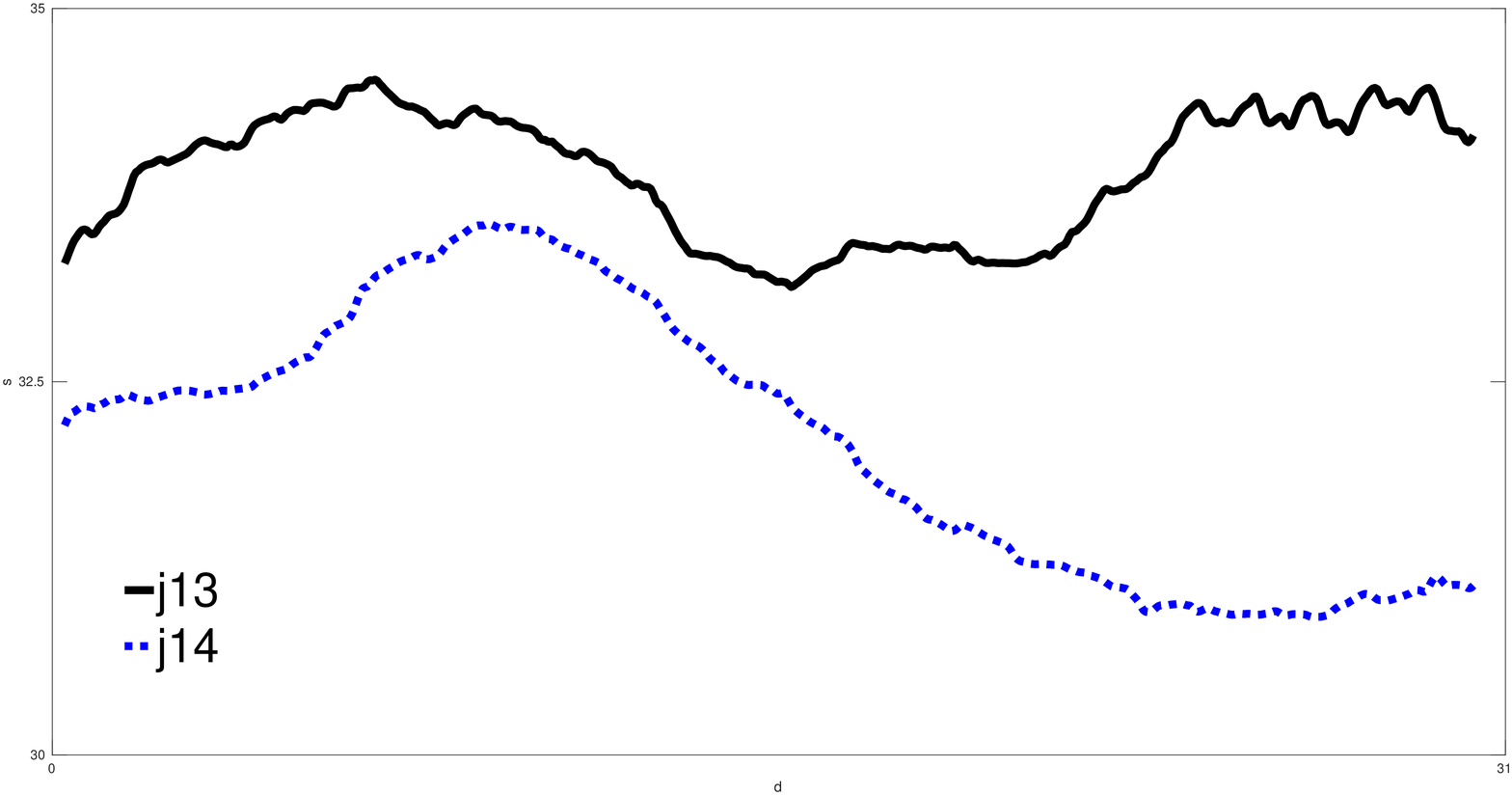}
                    \caption{January 2013 versus January 2014}
                    \label{jan1314}
                \end{subfigure}%
                \begin{subfigure}{0.33\textwidth}
                    \psfrag{d}{\hspace{0mm}\raisebox{-2mm}{\footnotesize{Day}}}
       				\psfrag{s}{\hspace{-3mm}\raisebox{6mm}{\footnotesize{km/h}}}
      				\psfrag{f13}{\hspace{0mm} \raisebox{0mm}{\scriptsize{Feb 2013}}}
      				\psfrag{f14}{\hspace{0mm} \raisebox{-0.5mm}{\scriptsize{Feb 2014}}}
      				\psfrag{0}{\hspace{0mm} \raisebox{-2mm}{\footnotesize{0}}}
      				\psfrag{29}{\hspace{-3mm} \raisebox{-2mm}{\footnotesize{29}}}
      				\psfrag{30}{\hspace{-5mm} {\footnotesize{30}}}
      				\psfrag{32.5}{\hspace{-6mm} {\footnotesize{32.5}}}
      				\psfrag{35}{\hspace{-5mm} \raisebox{-2mm}{\footnotesize{35}}}
					\includegraphics[width=\linewidth]{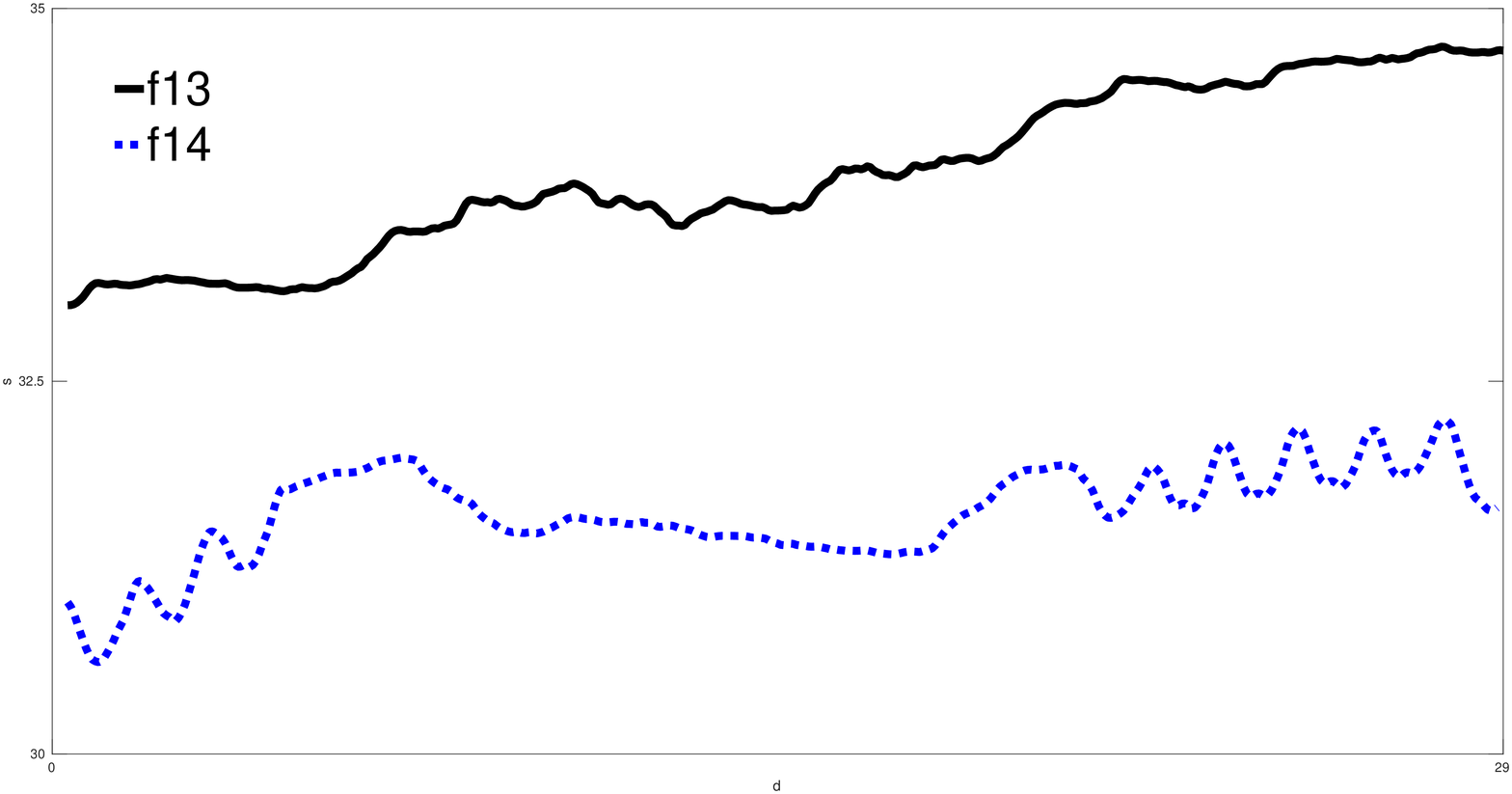}
                    \caption{February 2013 versus February 2014}
                    \label{feb1314}               
                \end{subfigure}%
                \begin{subfigure}{0.33\textwidth}
                    \psfrag{d}{\hspace{0mm}\raisebox{-2mm}{\footnotesize{Day}}}
       				\psfrag{s}{\hspace{-3mm}\raisebox{6mm}{\footnotesize{km/h}}}
      				\psfrag{m13}{\hspace{0mm} \raisebox{0mm}{\scriptsize{Mar 2013}}}
      				\psfrag{m14}{\hspace{0mm} \raisebox{-0.25mm}{\scriptsize{Mar 2014}}}
      				\psfrag{0}{\hspace{0mm} \raisebox{-2mm}{\footnotesize{0}}}
      				\psfrag{31}{\hspace{-3mm} \raisebox{-2mm}{\footnotesize{31}}}
      				\psfrag{31x}{\hspace{-5mm} {\footnotesize{31}}}
      				\psfrag{33.5}{\hspace{-6mm} {\footnotesize{33.5}}}
      				\psfrag{36}{\hspace{-5mm} \raisebox{-2mm}{\footnotesize{36}}}
					\includegraphics[width=\linewidth]{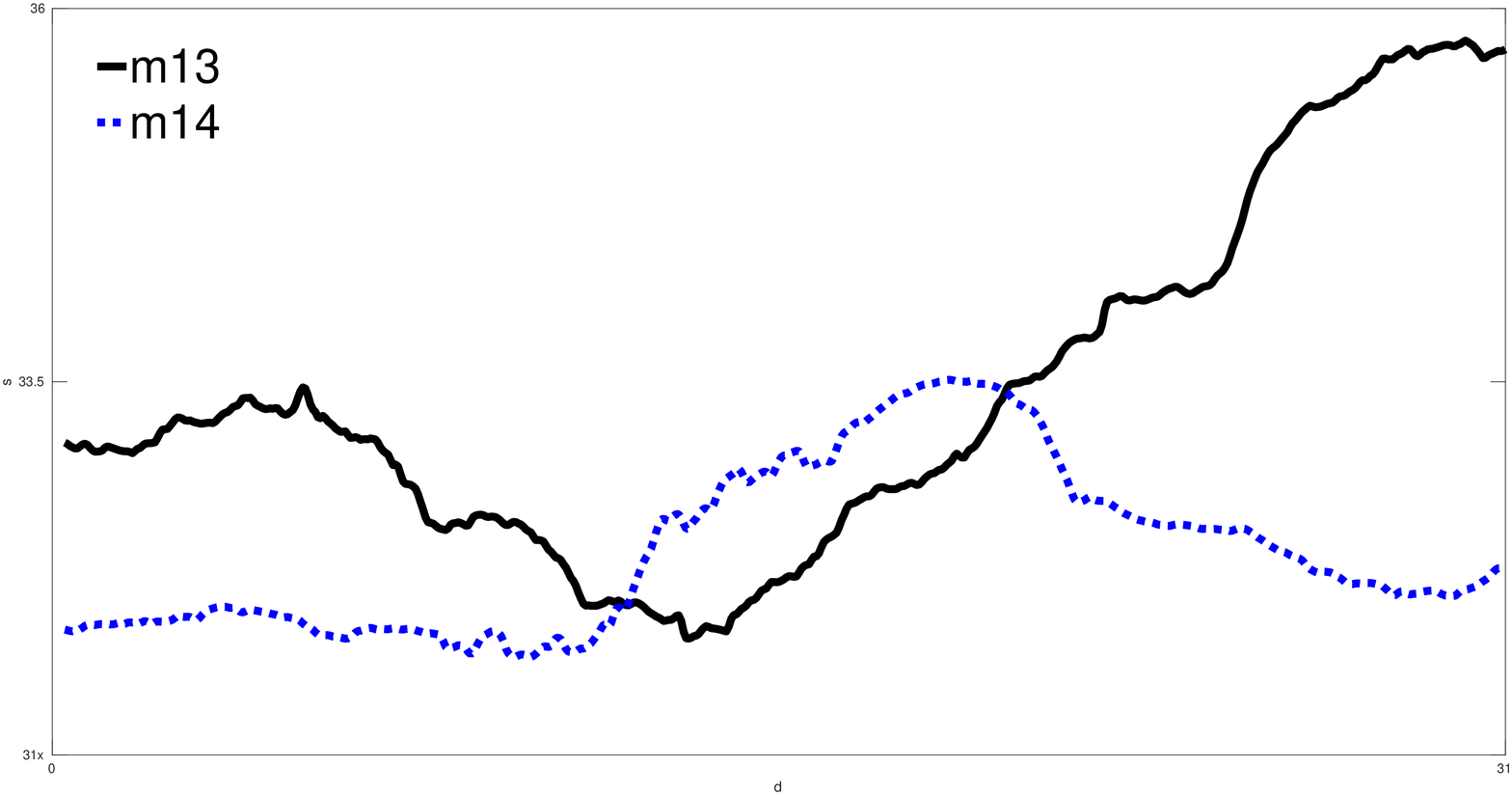}
                    \caption{March 2013 versus March 2014}
                    \label{mar1314}      
                \end{subfigure}%
                \caption{Comparison of average taxi speed in the first 3 months of 2013 and 2014. There is a visible reduction in the speeds in 2014 as compared to 2013.}
                \label{2013vs2014}
            \end{figure*}

The main contributions of this paper are twofold. First, by analysing the average speed of GPS enabled taxis over a period of one year, we provide statistical evidence for the increasing congestion levels in New Delhi. Second, we quantify the macroeconomic cost of traffic congestion in New Delhi, due to a variety of factors. \footnote{The conference version of this paper has appeared in the proceedings of the 2015 7th International Conference on Communication Systems and Networks (COMSNETS), held at Bangaluru, India in the year 2015. }

We begin with an analysis of the GPS traces from taxis to empirically show a downward trend in the average vehicular speed over the year 2013. Specifically, we employ a statistical test, known as the Kolmogorov–Smirnov test, which indicates that the average vehicular speed is statistically lower in the first quarter of 2014, as compared to the first quarter of 2013. We posit that this reduction in average speed of taxis is primarily due to an increase in congestion levels over the year of study. The major detrimental effects arising due to increasing congestion levels include the following.
\begin{itemize}
\item  People spend more time in traffic, leading to productivity losses.
\item  Vehicles spend more time idling, releasing more pollutants into the air.
\item  Increasing fuel wastages due to frequent traffic jams, and stalling at signals.
\end{itemize} 
In order to quantify these losses, we conduct a macroeconomic analysis of road congestion. To that end, we first aim to understand the marginal external costs of congestion, which measures the additional cost incurred due to an additional PCU worth of traffic.
This is estimated for three factors; namely, productivity losses,
air pollution costs, and road accidents. We have also included some key inferences obtained by analysing the major air pollutants on the roads of New Delhi. Next, we derive estimates for
the total costs of congestion. In addition to the previous factors
considered, we also incorporate fuel wastage due to traffic delays
in our computations. 

Once the congestion costs are estimated, it is then reasonable to consider some cost projections
based on historical trends. To that end, we project costs due to productivity
losses and air pollution till the year 2030. A key finding of our
study is to identify that productivity losses incurred by bus commuters
is the main contributing factor. This finding, coupled with the expected
increase in fuel wastage, highlights the need for a combination of
government policy and technology adoption. This work is an extended version of \cite{raymond}. In addition to the work presented in \cite{raymond}, we have analysed taxi traces for the city of Delhi to provide evidence for the rising congestion levels. The marginal and the total costs are also elaborated on in this paper.

The rest of the paper is organised as follows. In Section $\textrm{\textrm{II}}$, the motivation for estimating the economic costs of congestion is provided, by analysing data. In Section $\textrm{\textrm{III}}$ we compute the marginal costs of congestion, followed by Section $\textrm{\textrm{IV}}$,
in which we compute the total costs of congestion. In Section $\textrm{\textrm{V}}$,
we make projections on some of these costs based on vehicular growth
projections. Finally, in Section $\textrm{\textrm{VI}},$ we present
our conclusions and a few recommendations.

\begin{figure*}[h]
                \begin{subfigure}{0.33\textwidth}
                \psfrag{x}{\hspace{-2mm} \raisebox{-3mm}{\footnotesize{km/h}}}
       			\psfrag{y}{\hspace{-5mm}\raisebox{1.5mm}{\footnotesize{Taxi count}}}
     		 	\psfrag{j13}{\hspace{-2mm} \scriptsize{Jan 2013}}
      			\psfrag{j14}{\hspace{-2mm} \raisebox{-1mm}{\scriptsize{Jan 2014}}} 
     		    \psfrag{0}[][][1][-90]{\hspace{-2mm} {\footnotesize{0}}}
      			\psfrag{250}[][][1][-90]{\hspace{-2mm} {\footnotesize{250}}}
      			\psfrag{500}[][][1][-90]{\hspace{-2mm} \raisebox{-3.5mm}{\footnotesize{500}}}
      			\psfrag{2}{\hspace{-2mm} \raisebox{-2mm}{\footnotesize{< 20}}}
      			\psfrag{5}{\hspace{-2mm} \raisebox{-2mm}{\footnotesize{20-30}}}
      			\psfrag{8}{\hspace{-2mm} \raisebox{-2mm}{\footnotesize{30-40}}}
      			\psfrag{11}{\hspace{-2mm} \raisebox{-2mm}{\footnotesize{> 40}}}
      		     \includegraphics[height = 1.3in, width=\linewidth]{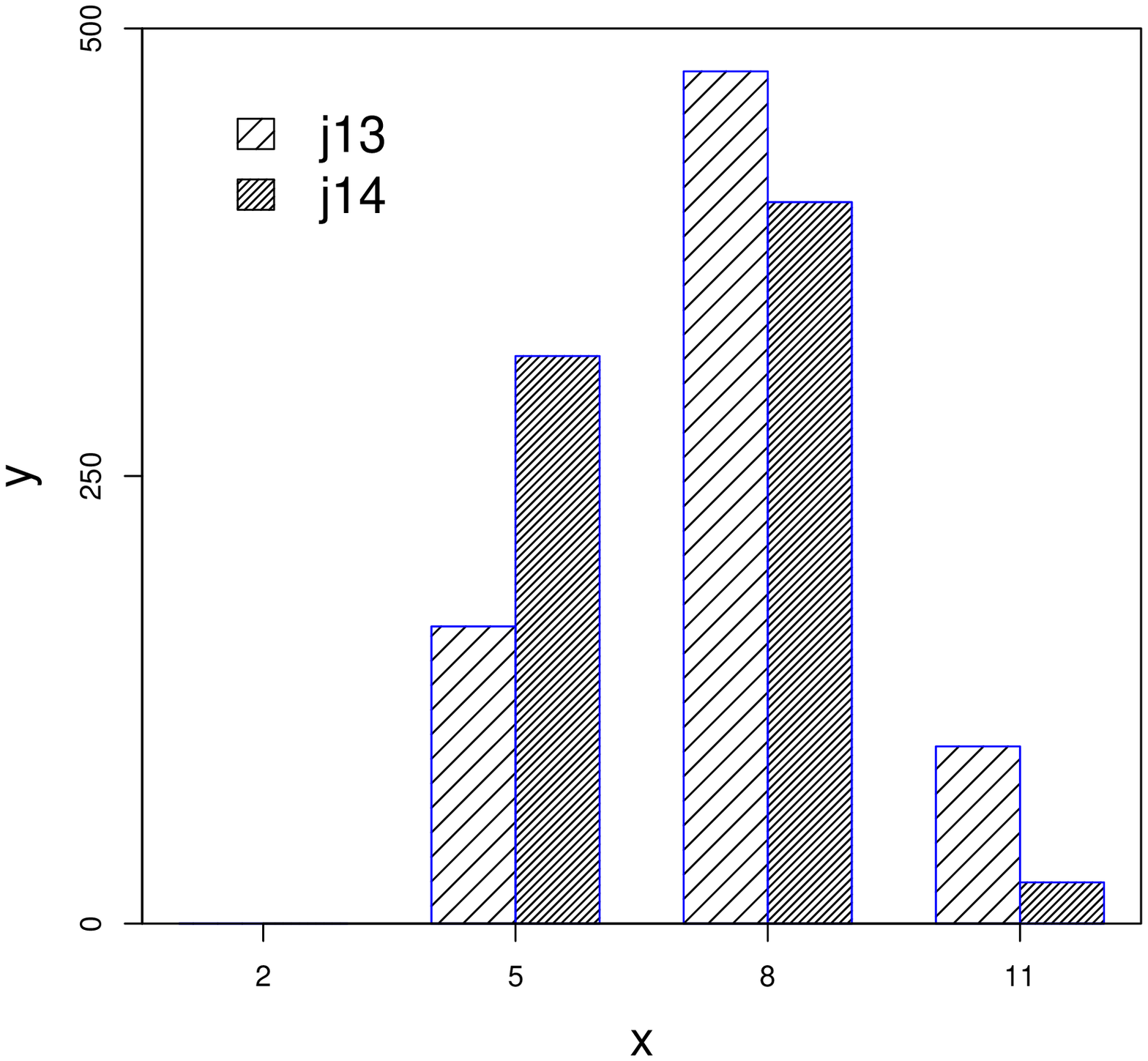}
                 \vspace{0.25mm}
                 \caption{January 2013 versus January 2014}
                 \label{jan1314_vehcount}
                \end{subfigure}%
                \begin{subfigure}{0.33\textwidth}
                \psfrag{x}{\hspace{-2mm} \raisebox{-3mm}{\footnotesize{km/h}}}
       			\psfrag{y}{\hspace{-5mm}\raisebox{1.5mm}{\footnotesize{Taxi count}}}
     		 	\psfrag{f13}{\hspace{-2mm} \scriptsize{Feb 2013}}
      			\psfrag{f14}{\hspace{-2mm} \raisebox{-1mm}{\scriptsize{Feb 2014}}} 
     		    \psfrag{0}[][][1][-90]{\hspace{-2mm} {\footnotesize{0}}}
      			\psfrag{250}[][][1][-90]{\hspace{-2mm} {\footnotesize{250}}}
      			\psfrag{500}[][][1][-90]{\hspace{-2mm} \raisebox{-3.5mm}{\footnotesize{500}}}
      			\psfrag{2}{\hspace{-2mm} \raisebox{-2mm}{\footnotesize{< 20}}}
      			\psfrag{5}{\hspace{-2mm} \raisebox{-2mm}{\footnotesize{20-30}}}
      			\psfrag{8}{\hspace{-2mm} \raisebox{-2mm}{\footnotesize{30-40}}}
      			\psfrag{11}{\hspace{-2mm} \raisebox{-2mm}{\footnotesize{> 40}}}
      		     \includegraphics[height = 1.3in,width=\linewidth]{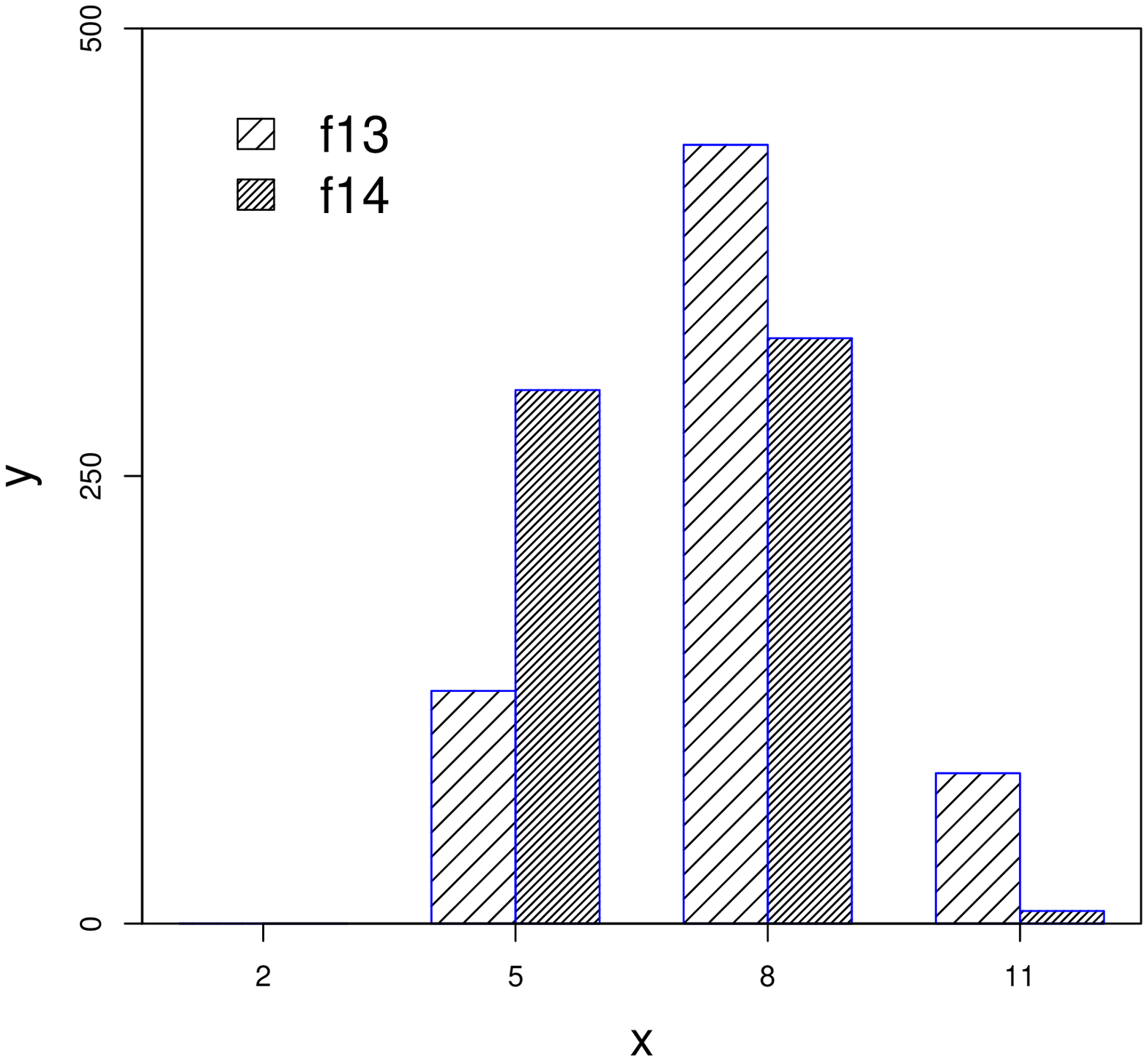}
                 \vspace{0.25mm}
                 \caption{February 2013 versus February 2014}
                 \label{feb1314_vehcount}       
                \end{subfigure}%
                \begin{subfigure}{0.33\textwidth}
                    \psfrag{x}{\hspace{-2mm} \raisebox{-3mm}{\footnotesize{km/h}}}
       			\psfrag{y}{\hspace{-5mm}\raisebox{1.5mm}{\footnotesize{Taxi count}}}
     		 	\psfrag{m13}{\hspace{-2mm} \scriptsize{Mar 2013}}
      			\psfrag{m14}{\hspace{-2mm} \raisebox{-1mm}{\scriptsize{Mar 2014}}} 
     		    \psfrag{0}[][][1][-90]{\hspace{-2mm} {\footnotesize{0}}}
      			\psfrag{250}[][][1][-90]{\hspace{-2mm} {\footnotesize{250}}}
      			\psfrag{500}[][][1][-90]{\hspace{-2mm} \raisebox{-3.5mm}{\footnotesize{500}}}
      			\psfrag{2}{\hspace{-2mm} \raisebox{-2mm}{\footnotesize{< 20}}}
      			\psfrag{5}{\hspace{-2mm} \raisebox{-2mm}{\footnotesize{20-30}}}
      			\psfrag{8}{\hspace{-2mm} \raisebox{-2mm}{\footnotesize{30-40}}}
      			\psfrag{11}{\hspace{-2mm} \raisebox{-2mm}{\footnotesize{> 40}}}
      		     \includegraphics[height = 1.3in,width=\linewidth]{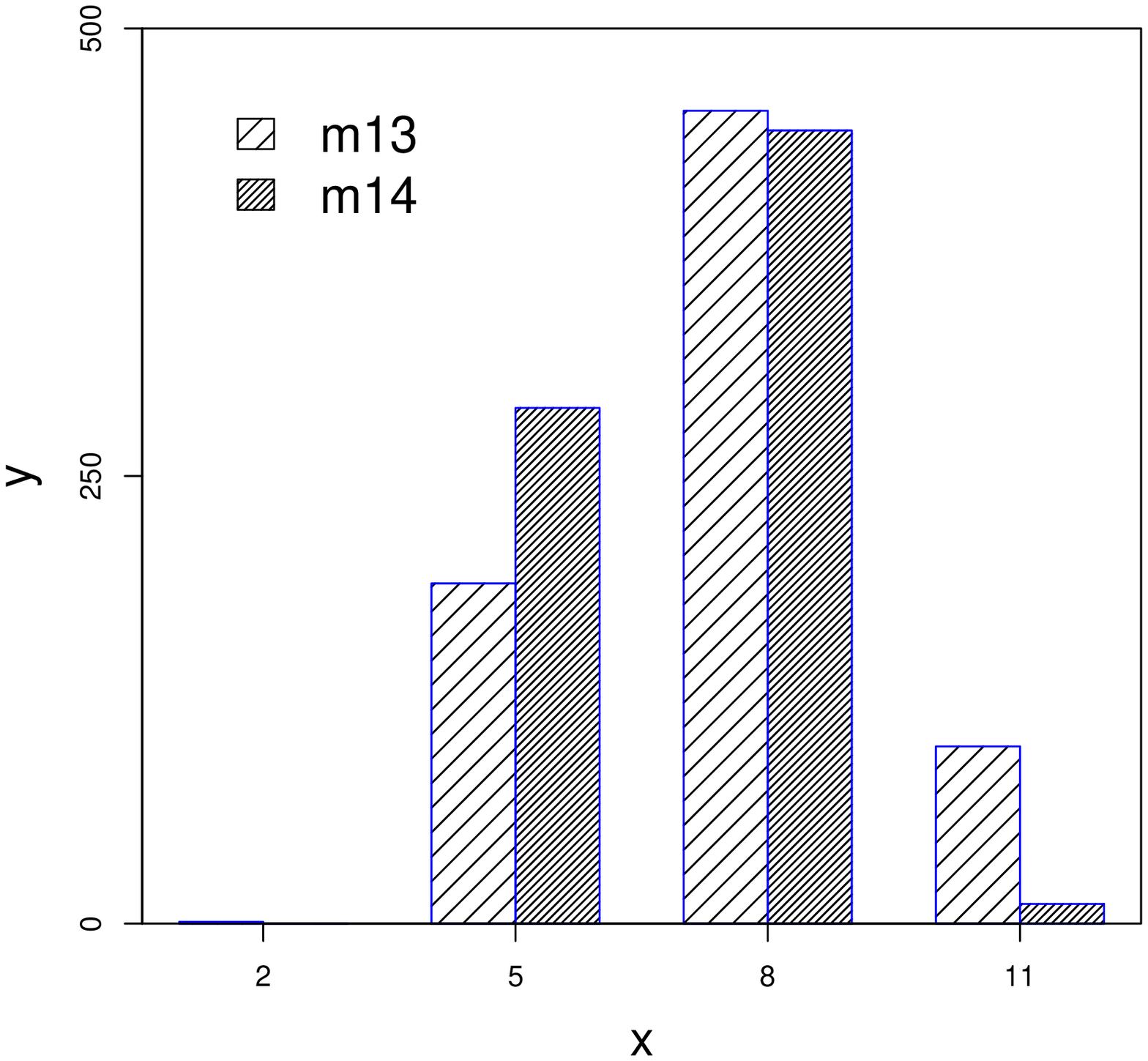}
                 \vspace{0.25mm}
                 \caption{March 2013 versus March 2014}
                 \label{mar1314_vehcount}
                \end{subfigure}%
                \caption{Number of taxis in various ranges of speed for the first 3 months of 2013 and 2014. There is a pronounced shift towards the lower speed ranges in 2014 compared to 2013.}
                \label{2013vs2014_vh}
            \end{figure*}

\section{Analysis of taxi traces}
In this section, we provide statistical evidence for the rising congestion levels. In particular, we perform an analysis of vehicle speed, using taxi GPS traces. The GPS traces used here are provided by a leading mobile application based taxi service provider. The data contain the vehicular position in terms of latitude and longitude, the current speed, the taxi ID and the direction in which it is heading. These GPS traces are available for a period of over one year, from January 2013 to March 2014. 

A prominent indicator of congestion in any city is the variation in average speed of vehicles over time. Observations show that for 78\% of the time in January 2014, the taxis exhibited a lower speed compared to January 2013. Similarly for February 2014 and March 2014, the corresponding percentages were 86\% and 71\% respectively. In fact, in January 2013, around 22\% of time, the speed was higher than the speed in January 2014 by 10\%. See Table \ref{pertimes} for similar inferences regarding the three months of interest. The data was also plotted for visual clarity. A basic moving average filter was used to smoothen the data. Figure \ref{2013vs2014} compares the average smoothed speed of taxis over the first 3 months of 2013 and 2014, \emph{i.e}, over the months of January, February and March.  We observe that there is a visible reduction in speed in 2014, as compared to 2013. While this effect is clearly visible in Figures \ref{jan1314} and \ref{feb1314}, for a brief period in mid-March (see \ref{mar1314}), taxis travelled with better speeds in 2014. After aggregating the speed over each hour in the year of 2013, we used a linear regression model to fit the data. The fit resulted in a negative slope, indicating that the average speed reduced from 34.2 km/h to 33.6 km/h in the year 2013. When similar procedure was repeated for the number of taxis, we observed a fit with a positive slope. It shows an increase in the number of taxis by roughly 200 units. Even though not conclusive, we can safely assume that other vehicles such as two wheelers and buses will follow a qualitatively similar reduction in the average speed. The negative trend for average speed and the positive trend for the vehicle count suggest an increase in road congestion over the period of study. 

Over the year 2013, vehicular speed on weekends was slightly lower than the speed on weekdays. This observation suggests that the roads in New Delhi suffer from more traffic jams on weekends than on weekdays. Apart from monitoring the average speed of vehicles, a similar congestion indicator is the count of vehicles in different ranges of speed. In figure \ref{2013vs2014_vh}, we observe a shift in the count of vehicles towards lower speed ranges in 2014. This is evident in all the 3 months that we analysed. The number of vehicles having average speed $>$ 40 has reduced marginally in an year. In the year 2014, the number of vehicles in the lower speed range (20-30 km/h) has increased, and in the higher speed range (30-40 km/h) has decreased as compared to 2013.

\begin{table}
\centering{}%
\begin{tabular}{|@{}c@{}||c|}
\hline 
Fraction of time (\%) & Reduction in speed (\%) \tabularnewline \hline \hline
\begin{tabular}{c|c|c}
Jan~ &  Feb~ & Mar
\end{tabular} 
&
\tabularnewline\hline
\begin{tabular}{c|c|c}
48.7 & 62.3 & 48.2
\end{tabular} 
&
>5\% 
\tabularnewline\hline
\begin{tabular}{c|c|c}
22.6 & 30.4 & 24.2
\end{tabular}
&
>10\% 
\tabularnewline\hline
\begin{tabular}{c|c|c}
03.1 & 03.6 & 04.2
\end{tabular} 
&
>20\%
\tabularnewline\hline
\begin{tabular}{c|c|c}
00.9 & 01.3 & 01.1
\end{tabular}
&
>25\%  \tabularnewline\hline
\end{tabular}
\caption{Percentage reduction in speed in 2014 compared to 2013}
\label{pertimes}
\end{table}

\subsection{Kolmogorov-Smirnov test}
In order to provide statistical evidence for the observations made by visual inspections, we conduct a statistical test known as the Kolmogorov-Smirnov test (K-S test) \cite{Massey51}. In the one sample variant of the K-S test, we test whether a specified test distribution could have generated the set of samples at hand. We are interested in the two-sample variant of the K-S test, which determines whether the two sets of samples differ significantly. The null hypothesis of the test states that the two sets of samples are drawn from the same distribution. We can reject the null hypothesis with  high confidence if its p-value is close to zero. On the other hand, a larger p-value indicates that the two sets of samples are statistically more similar. The K-S statistic $D$ captures the distance between the empirical distribution functions of two samples.

\begin{figure}
      \centering
     \psfrag{x}{\hspace{-4mm}\raisebox{-2mm}{\footnotesize{Speed}}}
      \psfrag{y}{\hspace{-2mm}\raisebox{1mm}{\footnotesize{CDF}}}

     \psfrag{s13}{\hspace{-2mm} \raisebox{-0mm}{\scriptsize{samples from 2013}}}
     \psfrag{s14}{\hspace{-2mm} \raisebox{-1mm}{\scriptsize{samples from 2014}}}  
      \psfrag{0.0}[][][1][90]{ \hspace{-1mm} \raisebox{-0mm}{\footnotesize{0}}}
      \psfrag{0.5}[][][1][-90]{\hspace{-1mm} \raisebox{-0mm}{\footnotesize{0.5}}}
      \psfrag{1.0}[][][1][-90]{\hspace{-1mm} \raisebox{-0mm}{\footnotesize{1}}}
       \psfrag{15}{\hspace{-2mm} \raisebox{-1mm}{\footnotesize{15}}}
      \psfrag{30}{\hspace{-2mm} \raisebox{-1mm}{\footnotesize{30}}}
      \psfrag{45}{\hspace{-2mm} \raisebox{-1mm}{\footnotesize{45}}}
      \includegraphics[height = 1.25in, width = 2.25in]{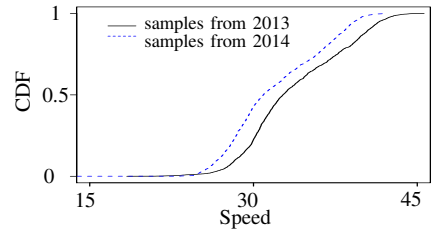}
      \vspace{0.25mm}      
      \caption{Empirical Cumulative Distribution function of samples from 2013 and 2014, where the population is the hourly taxi speed.}
      \label{ecdf_13_14}
\end{figure}

We run the K-S test for the speed data obtained from taxi GPS traces. The samples are drawn from the first 3 months of the years 2013 and 2014. We first consider the null hypothesis that the Cumulative Distribution Function (CDF) of samples from 2013 is equal to the distribution function of samples from 2014. When the K-S test was performed for this null hypothesis, it resulted in a p-value of 2.2e-15 and a $D$ statistic of 0.2088. This gives us overwhelming confidence to reject the null hypothesis, and suggests that the samples from 2013 and 2014 are statistically quite different. Similarly, we could reject with high confidence that the underlying distribution corresponding to the 2013 samples lies above that of the 2014 samples. Finally, we obtained a p-value greater than 0.9 for the hypothesis that the underlying distribution corresponding to the 2013 samples lies below that of the 2014 samples. This provides statistical evidence that the average speeds in 2013 were indeed greater than those in 2014. The empirical distribution of the average speeds, plotted in Figure~\ref{ecdf_13_14}, is also consistent with the findings of the K-S test.

%

Thus, the preliminary data analysis using taxi GPS traces suggest that there is a downward trend in the vehicular speed for the period we analysed. The statistical test further suggests that there is a high probability that the two speed distributions from 2013 and 2014 are dissimilar, and that 2013 has a higher speed distribution compared to 2014 for the same observation period. These inferences point to a visible increase in traffic congestion in 2014, when compared to 2013. 

We will now compute the losses due to congestion in terms of costs. For this, first, we calculate the marginal external costs of congestion and then, the total costs of congestion in the following sections.

\section{Marginal External Costs of Congestion}
The notion of marginal costs relates to the change in a dependent variable corresponding to a unit change in an underlying independent variable.
In the case of transportation systems, marginal costs of congestion refer to the costs incurred due to the addition of one vkm (vehicle kilometer) 
in an existing transportation network. Marginal costs indicate sensitivity of the transportation network to changes in demand. 
This in turn indicates the resilience of the transportation network \cite{carey}.  

An important distinction of marginal costs from similar measures is that marginal costs, in the case of 
road travel, almost always increase with the addition of a unit of demand. On the contrary, cost measures such as the average costs
may reduce with increased demand due to economies of scale, scope or density in the supply of transport services \cite{muller}. Marginal 
costs are of great practical importance especially in congestion pricing schemes that are gaining widespread acceptance in several cities around 
the world \cite{gris}. For instance, \cite{ozbay} emphasizes the importance of marginal costs due to the close estimation 
of real transportation costs accrued.

Several approaches have been followed to compute the marginal costs of congestion. One of the earliest works, \cite{walter}, computes the 
marginal costs by multiplying the per unit cost with the elasticity of the unit cost increased by one. The work makes use of traffic flow 
and velocity data collected by highway engineers, and lays emphasis on highway congestion. An important development in the computation of marginal 
costs is the inclusion of peak and off-peak loading as in \cite{glai}. Another important milestone in the study of marginal costs is \cite{newbery} - it brings into 
purview the costs due to road damage and the subsequent increased costs due to vehicles operating on these damaged roads. A more 
recent work in the area of marginal costs is \cite{link}, in which the authors carry out an extensive study of the components making up marginal 
costs and their implications for policy purposes. The work closest to this section is \cite{sen}, which computes the marginal costs in New Delhi with 
an elaborate methodology. However the work leaves out two important effects: two-wheelers and the effect of new legislation in New Delhi that has 
considerably reduced marginal costs due to air pollution \cite{foster}.

To compute marginal costs, the first step is to understand the different components that contribute to the marginal costs. 
A non-exhaustive list of components considered so far is given in Table \ref{table2}. The second step in computing marginal costs is to identify the components that are actually relevant. Ascertaining relevance of the components 
includes considering the geographical, legal and regional particulars, unique to the transportation network. For the purpose of computing the marginal costs in New Delhi, we
consider only the following three components: productivity losses, environmental costs and accidents. The other costs are neglected for the following reasons:

\begin{itemize}
\item Infrastructure and maintenance costs are marginal in the case of New Delhi. Due to the developing nature of the
Indian economy skilled labour is relatively cheaper \cite{taylor}, and hence implies reduced infrastructure costs. 

\item  Operation and usage costs of vehicles, have been increasing globally. In the case of India, the effect of 
increasing operation and usage costs has been very gradual due to the increasing quality of infrastructure, and customer-facing technology \cite{iyer}. 

\item Additional service costs and Mohring effect can be neglected in the case of New Delhi, since state 
transportation schedules are not dynamic, and often times do not reflect the demand. Despite several 
studies highlighting the importance of a dynamic scheduling system for Indian cities, progress in its implementation is scarce  and the schedules are more or less fixed \cite{verma}.

\item Fuel wastage is a relevant component for computing marginal costs in New Delhi. However, due
to restricted availability of information, fuel wastage can only be considered as a component 
while computing total costs of congestion. It will be taken up in the next section.
\end{itemize}
\begin{table}
\centering{}%
\begin{tabular}{|c||c|}
\hline 
Category  & Basis \tabularnewline
\hline 
\hline 
Infrastructure maintenance  & Maintenance costs due to road usage \tabularnewline
\hline 
Operation and usage & Cost of an additional vkm \tabularnewline
\hline 
Productivity losses  & Cost due to delays  \tabularnewline
\hline 
Additional service  &  Cost due to providing  remedial services \tabularnewline
\hline 
Mohring effect  & Benefits due to increased demand\tabularnewline
\hline 
Accidents  & Expected increase due to additional road travel \tabularnewline
\hline 
  Emission and pollution & Increased noise and emission costs  \tabularnewline
\hline 
Fuel wastage  & Increased consumption costs \tabularnewline
\hline

\end{tabular}
\caption{Components making up marginal costs \cite{muller} }
\label{table2}
\end{table}

The third step, is to examine the selected components individually and compute the marginal costs due to each of these components. 
The computational approach is similar to that found in \cite{sen}, but with a few important modifications. The modifications include 
considering two-wheelers and taking into account the post legislation CNG bus policy. 

From a total costs perspective, if $T_i$ is the total cost of congestion function, due to the $i^{th}$ contributing component. $\nu_C$, $\nu_f$ are the average 
vehicle speeds under congested and free-flow conditions respectively. Then, a plausible form for $T_i$ is:

\begin{equation}
T_i = F_i(\nu_C) - F_i(\nu_f). \label{eq:total-form}
\end{equation}

Here, $F_i(\nu)$ is a function that represents total costs due to the $i^{th}$ contributing component when the average 
network speed is $\nu$. The derivative of total costs with respect to vkm is expected to directly give us the marginal costs. Note that 
the derivative of the second term in Equation (\ref{eq:total-form}) vanishes or leaves a very small 
contribution, depending on the network characteristics, since the
average free-flow speed is constant. Hence, the marginal costs due to air pollution and 
accidents may be approximated to be the marginal costs of congestion due to air pollution and accidents. The following subsections 
compute the marginal costs due to each of the corresponding components. In the end, the total marginal costs due to all these contributions  are computed.

\subsection{Productivity Losses}

Productivity losses are costs incurred due to delays experienced by commuters. The losses can be categorized into two dimensions:

\begin{itemize}
 
\item A personal dimension that covers losses arising out of  personal time forgone while stuck in traffic delays. It includes time, that could
be used towards employment, rest or any personally gainful activity.  

\item A commercial dimension, especially in the freight and cargo industry. Productivity 
losses may stem out of canceled orders or refused shipments due to late delivery. 

\end{itemize}

While the first aspect that covers productivity losses has been widely studied, fewer works have taken up business impact
caused by congestion \cite{ciecs}. We will stick with losses on a personal scale, since business impact of traffic congestion 
is difficult to be modeled for New Delhi. 

To compute the productivity losses for an additional vkm, the first step is to
specify a speed-flow relationship for a given mode, $i$ at time $j$.
We use the Passenger Car Units (PCU) metric in the speed-flow relationship as used in most works to capture 
the vehicle characteristics \cite{sen}. PCU is the impact that a transport mode has on traffic variables such as speed and it is compared against a car. For example, a motorcycle is considered as 0.5 PCU. A 
difference in our methodology is that we use PCUs for two-wheelers 
on Indian roads as investigated in \cite{chandra}. 
The commonly used speed-flow relationship is given by:

\begin{equation}
t_{ij}=A_{1j}\big[A_{2}+A_{3}\exp(A_{4}q_{i})\big].\label{eq:speed-flow}
\end{equation}

Here, $t_{ij}$ is the time (in minutes) needed to
travel $1$ km on mode $i$ during time interval $j$. $A_{1j},$
$A_{2}$, $A_{3}$ and $A_{4}$ are constants that depend on the characteristics of the transportation 
network under consideration. $A_{1j}$ also depends on the period of travel
$j$, and $q_{i}$ represents the PCU of mode $i$. Note that this approach is justified 
at least in the case of New Delhi, since the speed-flow fit has a high $R^2$ measure \cite{sen}.

The Marginal 
Economic Costs of Congestion
due to Productivity loss ($\text{MECCP}_{i}$) is thus given by:
\begin{equation}
\text{MECCP}_{i}=\mathlarger{\sum}_{j}\frac{\partial t_{ij}}{\partial q_{i}}x_{ij}\text{VOT}_{ij}.
\end{equation}

In the above equation, $x_{ij}$ is the number of passenger
kilometers (pkm) travelled in period $j$ by mode $i$. $\text{VOT}_{ij}$ is the 
Value Of Time for a user travelling
in mode $i$, during period $j$.
$\frac{\partial t_{ij}}{\partial q_{i}}$ is the increase in delay
suffered in mode $i$ during period $j,$ due to a unit increase in
the PCU of mode $i.$ Across modes, the value of time for commuters estimated in \cite{sen} is
corrected to reflect present day price-levels by using:

\begin{equation}
P_{new}=P_{old}\frac{\Gamma_{2013}}{\Gamma_{2001}}\frac{1}{[1+\frac{i_{2013}-i_{2001}}{100}]}, \label{eq:inflation}
\end{equation}

where, $P_{old}$ is the original 2001 price
(in Rupees per hour) used in \cite{sen} for the value of time for different
modes of transport. $\Gamma_{t}$ is the price-level in New Delhi during
the year $t$, and $i_{t}$ is the national inflation rate during
year $t$. A textbook definition of price-level describes it as the sum of the prevailing
prices of a standard basket of goods and services consumed indicating the prevailing value of money.
 Table \ref{table3}
lists the value of time for passengers using different modes of transport.
Comparing 2001 to 2013 levels (Table \ref{table4}), a near doubling in the marginal 
costs due to productivity is observed.

\begin{table}
\centering{}%
\begin{tabular}{|c||c|c|}
\hline 
Mode of Transport  & 2001 (in INR/h)  & 2013 (in INR/h)\tabularnewline
\hline 
\hline 
Car  & 50.84  & 99.12\tabularnewline
\hline 
Bus  & 17.11  & 33.36\tabularnewline
\hline 
Two-wheeler  & 25.74  & 50.18\tabularnewline
\hline 
\end{tabular}
\caption{Value of time for passenger transport \cite{sen}}
\label{table3}
\end{table}

\begin{table}
\centering{}%
\begin{tabular}{|c||c|c|}
\hline 
Mode of Transport  & 2001 (inINR/vkm)  & 2013 (inINR/vkm)\tabularnewline
\hline 
\hline 
Car  & 4.91  & 9.57\tabularnewline
\hline 
Bus  & 9.83  & 19.16\tabularnewline
\hline 
Two-wheeler  & 0.98  & 1.91\tabularnewline
\hline 
\end{tabular}
\caption{Marginal costs of congestion due to productivity losses }
\label{table4}
\end{table}

\subsection{Air Pollution Costs}
Vehicular emissions cause serious 
air pollution problems and are a health hazard. Air pollution costs arise from health and environmental damages due to vehicular emissions. Increased traffic congestion stalls vehicles 
and increases on-road time, which in turn considerably increases vehicular emissions. Computing marginal costs of congestion due to air pollution
entails considerations such as emission per vehicle kilometer (vkm),
vehicle fleet age structure, and the estimates of pollution costs
per unit of the pollutant. The Marginal External Costs of Congestion
due to Emissions ($\text{MECCE}_{i}$) for a transport mode $i,$
summed over all emitted pollutants indexed by $k$, is given by:

\begin{equation}
\text{MECCE}_{i}=\sum_{k}\rho_{i}^{k}\delta^{k}.\label{eq:MECCE}
\end{equation}

In the above equation, $\rho_{i}^{k}$ is the age-division-corrected
emission structure for the $i^{th}$ mode, and $\delta^{k}$ is the cost per kilogram of pollutant $k$ emitted,
computed in \cite{mandal}. Further, $\rho_{i}^{k}$ is computed using: 

\begin{equation}
\rho_{i}^{k}=\sum_{j}E_{ij}^{k}\gamma_{ij},
\end{equation}

where, $E_{ij}^{k}$ is the coefficient of emission per vkm for the $i^{th}$
mode of transport, belonging to the $j^{th}$ age-division, for the
$k^{th}$ pollutant. 

The emission coefficients are listed in Tables
\ref{table5}, \ref{table6} and \ref{table7}. Though vehicular emissions consist of a large variety of GHGs (Green House Gases) as well as harmful pollutants, in this study we only consider pollutants that are emitted in significant amounts, such as: carbon monoxide (CO), hydrocarbons (HC), oxides of nitrogen $(\text{NO}_{X})$,
and Particulate Matter (PM). In Table \ref{table6} two
differing sets of emission coefficients for PM are provided. The `Business As Usual'
(BAU) type corresponds to the PM emission coefficients, if the ruling
to convert all buses to Compressed Natural Gas (CNG) had not been enforced
in New Delhi. The next column in the table provides the most recent PM emission coefficients, following the ruling. Though the vehicle fleet-age structures differ, the PM emission coefficients remain the same, since most older vehicles have been refit to comply with the existing standards.

\begin{table}
\centering{}%
\begin{tabular}{|c||c|c|c|c|}
\hline 
Age Group  & CO  & HC  & $\text{NO}_{X}$  & PM\tabularnewline
\hline 
\hline 
1991-1995  & 4.75  & 0.84  & 0.95  & 0.06\tabularnewline
\hline 
1996-2000  & 4.53  & 0.66  & 0.75  & 0.06\tabularnewline
\hline 
2001-2005  & 3.01  & 0.19  & 0.12  & 0.05\tabularnewline
\hline 
2006-2010  & 0.84  & 0.12  & 0.09  & 0.03\tabularnewline
\hline 
\end{tabular}
\caption{Emission coefficient $E_{ij}^{k}$ for cars in New Delhi (in gm/km) \cite{sen,arai} }
\label{table5}
\end{table}

\begin{table}
\begin{centering}

\par\end{centering}

\centering{}%
\begin{tabular}{|c||c|c|c|c|c|}
\hline 
Age Group  & CO  & HC  & $\text{NO}_{X}$  & PM (BAU)  & PM (Actual)\tabularnewline
\hline 
\hline 
1991-1995  & 13.06  & 2.40  & 11.24  & 2.013  & 0.032\tabularnewline
\hline 
1996-2000  & 4.48  & 1.46  & 15.25  & 1.213  & 0.032\tabularnewline
\hline 
2001-2005  & 3.97  & 0.26  & 6.77  & 1.075  & 0.032\tabularnewline
\hline 
2006-2010  & 3.92  & 0.16  & 6.53  & 0.300  & 0.032\tabularnewline
\hline 
\end{tabular}
\caption{Emission coefficient $E_{ij}^{k}$ for buses in New Delhi (in gm/km)
\cite{arai,santosh} }
\label{table6}
\end{table}

\begin{table}
\begin{centering}

\par\end{centering}

\centering{}%
\begin{tabular}{|c||c|c|c|c|}
\hline 
Age Group  & CO  & HC  & $\text{NO}_{X}$  & PM\tabularnewline
\hline 
\hline 
1991-1995  & 3.12  & 0.78  & 0.23  & 0.010\tabularnewline
\hline 
1996-2000  & 1.58  & 0.74  & 0.30  & 0.015\tabularnewline
\hline 
2001-2005  & 1.65  & 0.61  & 0.27  & 0.035\tabularnewline
\hline 
2006-2010  & 0.72  & 0.52  & 0.15  & 0.013\tabularnewline
\hline 
\end{tabular}
\caption{Emission coefficient $E_{ij}^{k}$ for two-wheelers in New Delhi (in gm/km) \cite{arai}}
\label{table7}
\end{table}

\begin{table}
\begin{centering}

\par\end{centering}

\centering{}%
\begin{tabular}{|c||c|c|}
\hline 
Pollutant  & High Estimate & Low Estimate\tabularnewline
\hline 
\hline 
CO  &  0.46 & 0.05\tabularnewline
\hline 
HC  & 6.73 & 0.60\tabularnewline
\hline 
$\text{NO}_{X}$  &  108.26 & 7.37 \tabularnewline
\hline 
PM  & 869.57 & 63.73\tabularnewline
\hline 
\end{tabular}
\caption{High and Low cost estimates of $\delta^k$ (in INR/kg) for New Delhi \cite{mandal}}
\label{table8}
\end{table}

The coefficient $\gamma_{ij}$, representing the distribution of vehicle
fleet age structure, is given in Table \ref{table9} for
New Delhi. Hence, using (\ref{eq:inflation}), (\ref{eq:MECCE}) and $(6)$,
the marginal external costs of congestion due to air pollution can
be readily computed; see Table \ref{table10}.

We note that even for the newer buses from the 2006-2010 age group,
there is an order of magnitude difference between the actual and BAU
values for PM emission coefficients. For the older buses, the improvement
in PM emissions can be as large as two orders of magnitude. This highlights
the pivotal role and potential of government policy and enforcement in this area.

\begin{table}
\centering{}%
\begin{tabular}{|c||c|c|c|}
\hline 
Age Group  & Cars  & Buses  & Two-wheelers\tabularnewline
\hline 
\hline 
1991-1995  & 0.154  & 0.374  & 0.398\tabularnewline
\hline 
1996-2000  & 0.200  & 0.593  & 0.252\tabularnewline
\hline 
2001-2005  & 0.323  & 0.016  & 0.235\tabularnewline
\hline 
2006-2010  & 0.323  & 0.016  & 0.115\tabularnewline
\hline 
\end{tabular}
\caption{Vehicle fleet age structure $\gamma_{ij}$ for vehicles operating
in New Delhi \cite{arai}}
\label{table9}
\end{table}

\begin{table}

\centering{}%
\begin{tabular}{|c||c|}
\hline 
Mode of Transport  & Cost (INR/vkm)\tabularnewline
\hline 
\hline 
Car  & 0.26\tabularnewline
\hline 
Bus  & 1.78\tabularnewline
\hline 
Two-wheeler  & 0.12\tabularnewline
\hline 
\end{tabular}
\caption{Marginal external costs of congestion due to air pollution}
\label{table10}
\end{table}

\subsection{Accidents}

Accident costs arise mainly from factors such as manpower
losses, vehicular damages, insurance and other exigency costs. Accident statistics for the year 2013 is as given in Table \ref{table11}.
The economic value of damages due to accidents has been assessed in
\cite{tcs}. In computing the cost due to accidents, the same prices are used, but these prices are corrected to 2013 levels using (\ref{eq:inflation}). The corrected
prices are given in Table \ref{table12}. The Marginal Economic Costs of Congestion due to Accidents involving the $ i^{th}$ mode ($\text{MECCA}_i$) is given by:

\begin{equation}
\text{MECCA}_i = \frac{\sum_l{\epsilon_{il}}H_{l}}{365\psi_i\mu_i}.
\end{equation}


In the above equation, $\epsilon_{il}$ is the number of accidents in the $l^{th}$ seriousness category
for the $i^{th}$ mode as in Table \ref{table11}. Table \ref{table12} gives $\text{H}_{l}$, the average cost 
of the accident corresponding to the particular mode and the seriousness category. $\Psi_i$ is the average total number of 
trips in a day as given in Table \ref{table18} and $\mu_i$  is the average length of a 
trip for mode $i$ as in \cite{rites}. Following this computation, the marginal 
costs due to accidents are tabulated in Table \ref{table14}.

\begin{table}

\centering{}%
\begin{tabular}{|c||c|c|c|c|}
\hline 
Accident Classification  & Events  & Bus  & Cars  & Two-wheelers\tabularnewline
\hline 
\hline 
Minor Accidents  & 169  & 25  & 44  & 27\tabularnewline
\hline 
Major Injury Accidents  & 5619  & 843  & 1461  & 899\tabularnewline
\hline 
Fatal Accidents  & 1778  & 338  & 213  & 124\tabularnewline
\hline 
Persons Injured  & 7098  & N / A  & N / A  & N / A\tabularnewline
\hline 
Persons Killed  & 1820  & N / A  & N / A  & N / A\tabularnewline
\hline 
\end{tabular}
\caption{Accident statistics for New Delhi, $\epsilon_{il}$ \cite{police}}
\label{table11}
\end{table}

\begin{table}
\centering{}%
\begin{tabular}{|c||c|}
\hline 
Accident Classification  & Cost (INR in 2013-14 prices)\tabularnewline
\hline 
\hline 
Fatality  & 1745600\tabularnewline
\hline 
Major Accident  & 311430\tabularnewline
\hline 
Minor Accident / Non Injury  & 40917\tabularnewline
\hline 
\end{tabular}
\caption{Economic costs of accidents, $\text{H}_{l}$ \cite{tcs}}
\label{table12}
\end{table}

\begin{table}
\begin{tabular}{|c||c|}
\hline 
Mode of Transport  &  Distance\tabularnewline
\hline 
\hline 
Bus  & 77571669\tabularnewline
\hline 
Car  & 32735914\tabularnewline
\hline 
Two-wheeler  & 32605073\tabularnewline
\hline 
\end{tabular}
\centering{}\caption{Average vkm perday commuted in New Delhi, $\psi_i\mu_i$ (in km/day) \cite{rites}}
\label{table13}
\end{table}

\begin{table}

\begin{tabular}{|c||c|}
\hline 
Mode of Transport  & Marginal Cost (in INR/vkm)\tabularnewline
\hline 
\hline 
Bus  & 1.578\tabularnewline
\hline 
Car  & 0.042\tabularnewline
\hline 
Two-wheeler  & 0.113\tabularnewline
\hline 
\end{tabular}
\centering{}\caption{Marginal external costs of congestion due to accidents}
\label{table14}
\end{table}

\begin{table}
\centering{}%
\begin{tabular}{|c||c|c|c|c|c|}
\hline 
Mode  & Lost Time  & Pollution  & Accidents  & Total  & Total \cite{sen} \tabularnewline
\hline 
\hline 
Bus  & 19.16  & 1.78  & 1.58  & 22.52  & 26.23\tabularnewline
\hline 
Car  & 9.57  & 0.26  & 0.04  & 9.87  & 6.29\tabularnewline
\hline 
Two-wheeler  & 1.91  & 0.12  & 0.11  & 2.14  & N / A\tabularnewline
\hline 
\end{tabular}
\caption{Marginal costs of congestion in New Delhi (in INR/vkm)}
\label{table15}
\end{table}

\subsection{Total Marginal Costs}

Total marginal costs of congestion due to the three factors considered
(productivity losses, air pollution, and accidents) are summed in
Table \ref{table15}. In Table \ref{table16}, the contribution due to the air pollution component to the
total marginal costs are compared against the results obtained by \cite{sen}. The key findings from these marginal cost estimates
are as follows: 
\begin{itemize}
\item The most significant increase in marginal costs is for cars, estimated
at nearly 57\%. In contrast, the corresponding figure for
buses is only 10.4\%. 
\item A striking observation is the decrease in the marginal costs
due to air pollution in 2013. The contribution of the air pollution
component has reduced despite the increased cost per gram of emissions
corrected to the 2010 prices. This appears to be a direct consequence
of government policy: (i) the switch to CNG buses, and (ii) the complete
phasing out of vehicles purchased before 1990. 
\end{itemize}


\begin{table}
\centering{}%
\begin{tabular}{|c||c|c|}
\hline 
Mode  & Marginal costs  & Marginal costs from \cite{sen}\tabularnewline
\hline 
\hline 
Car  & 0.26  & 0.27 - 2.74\tabularnewline
\hline 
Bus  & 1.78  & 9.12 - 14.14\tabularnewline
\hline 
Two-wheeler  & 0.12  & N / A\tabularnewline
\hline 
\end{tabular}
\caption{Air pollution component contribution to the total marginal costs of
congestion}
\label{table16}
\end{table}

%
%

\section{Total Costs of Congestion}

The total costs of congestion are the sum of all costs accrued due to the delays experienced 
arising out of stalled speeds caused by road traffic congestion. 
In most cases, the 
total costs of congestion are defined with respect to a baseline scenario where congestion is minimal. The excess
costs over and above the operating points of this scenario are  considered as the total costs of congestion \cite{muller}. This popular
approach highlights the dependence of total costs of congestion on not only the number of vehicles, but also the 
transportation network aspects such as capacity.

Table \ref{table17} provides estimates of total costs of congestion made 
by several works for several transportation networks around the world. One aspect is clear, the cost 
has been consistently rising. Note that the costs of congestion computed by these works, correspond to
the price levels when the research was actually published.

\begin{table}
\centering{}%
\begin{tabular}{|c||c|}
\hline 
Work & Estimate\tabularnewline
\hline 
\hline 
Glanville, 1958  & GBP 170 million in the UK \tabularnewline
\hline 
 Newbery, 1995 & GBP 19.1 billion in the UK \tabularnewline
\hline 
Dodgson and Lane, 1997 & GBP 7 billion in the UK\tabularnewline
\hline 
Mumford, 2000 & GBP 18 billion in the UK\tabularnewline
\hline 
Tweedle, et al., 2003 & GBP 24 billion in the UK\tabularnewline
\hline 
Scottish Executive, 2005 & GBP 71 million in 10 areas of Scotland\tabularnewline
\hline 
DoTRS - Canberra, 2007 & AUD 6.1 billion in Melbourne\tabularnewline
\hline
CEBR, 2014 & US\$ 20.5 billion in the UK\tabularnewline
\hline
CEBR, 2014 & US\$ 124 billion in the US\tabularnewline
\hline

\end{tabular}
\caption{Estimates of total congestion costs \cite{muller}}
\label{table17}
\end{table}

%
%
%
%
%
%
%
%

As we compute the total costs of congestion in the proceeding sections, it is also important to 
understand some issues regarding the total costs of congestion. Several authors have in 
the past questioned the meaning of the total costs of congestion. Some of the criticisms are:

\begin{itemize}

\item  The `total cost of congestion' is rather a misnomer. If the total costs of congestion are 
incurred due to congestion, does alleviating congestion guarantee that the economy will be better 
off by an amount equal to the total costs of congestion? Certainly not. Alleviating congestion implies
infrastructural spending, which has to be meted out by the state \cite{goodwin}.

\item Several paradoxes relating to transportation networks 
have proved that reducing congestion implies reducing travel impedance and hence 
increasing travel demand. Increased travel will only increase the total costs of congestion \cite{goodwin}. 

\item The total costs of congestion measures are also criticized because 
the baseline scenario is rather arbitrary. Accuracy of measures of 
average free-flow speeds have been questioned \cite{goodwin}.

\end{itemize}

Despite questions raised on the utility and accuracy of total costs of congestion, such 
a measure is important in the case of a developing country like India. Some of the reasons for this are:
\begin{itemize}
\item India does not have an established system 
of basic infrastructure. For instance, metro transportation is yet to be opened in most of the cities. Expenditure on these facilities 
will considerably debunk congestion, while in the case of developed countries with already existing transportation
facilities, increased expenditure may only marginally provide relief.

\item Total costs provide an excellent direction for a country like India which is still in the 
planning phases. Cities are still being built - not the case in developed countries.

\end{itemize}

Total costs of congestion have been well-studied in the past. Though there are several variations in the computation process, the basic
framework remains the same in all past works: compare the congestion scenario with a reference baseline scenario with minimal congestion. The earliest 
work, perhaps, on the total costs of congestion is \cite{glan}. The approach followed in the computation makes use of 
the basic delay aspect. However, \cite{glan} neglects the value of non-work time. In \cite{newbery}, the authors provide a 
different approach by categorizing road users and then computing a nationwide total cost figure for the UK. This approach has been 
criticized in \cite{dodgson} on dimensional counts, for multiplying marginal costs with a total volume. In \cite{dodgson}, the authors use a link-based 
methodology to estimate time and operating costs, and then compare costs at free-flow and current speeds. In our approach, we study 
total costs of congestion by aggregating costs for the three factors considered: productivity losses, 
air pollution costs and accidents. An additional cost considered in the case of total costs is the fuel wastage costs. All these 
costs are computed by comparing against a baseline scenario, mainly in terms of average speeds.

Though by definition a simple numerical integration of the marginal cost function seems to
intuitively provide the total costs of congestion, the computation
of the marginal cost function throughout the range of the integral
is cumbersome due to changes in parameters $A_{1j}$, $A_{2}$, $A_{3}$ and $A_{4}$ in $(2)$ as the number of vehicles change. Therefore, a data driven 
approach is adopted, that considers the prevailing
averages of the various parameters that have been considered in the
previous section.

The following subsections delve into the computational details of the total costs of congestion. We will first consider 
the contribution by productivity losses to the total costs, and followed by this, air pollution costs will be studied. Costs due to 
accidents follow, finally ending with the contribution of fuel wastage to the total costs.

\subsection{Productivity Losses}

Here, the productivity losses entail a total approach, \emph{i.e}, losses 
incurred by all commuters due to delays caused by all vehicles in the network. An important point worth 
mentioning here is that, the productivity losses in this case will depend on the average vehicle occupancy. The reason 
for dependency on occupancy is that the costs in this case are not with respect to a vehicular parameter (such as vkm), but
necessitate the inclusion of an aggregate passenger number to compute losses.

Computing the total costs of congestion due to productivity losses involves considering several factors. These include: Value 
of time, average occupancy, trip length by mode, number of trips by mode, free-flow speed and average speed in 
congested conditions. Computing most of these factors at an individual micro-level is a formidable task. So for the purpose of
these computations, averaged values of these factors are available, and are expected to produce similar results. Considering these factors, the Total
Costs of Congestion due to Productivity Losses (TCCPL) is then given by:

\begin{equation}
\text{TCCPL}=365\sum_{i}\text{VOT}_{i}\Psi_{i}\mu_{i}\Lambda_{i}\left(\frac{1}{\nu_{C}}-\frac{1}{\nu_{f}}\right).\label{eq:TCCPL}
\end{equation}

In the above equation, $\Psi_{i}$
is the average total number of trips in a day for mode $i$, $\mu_{i}$ (in
kilometres) is the average length of a trip for mode $i$, and $\Lambda_{i}$ is the average occupancy for mode $i$. $\text{VOT}_{i}$ (in Rupees per hour)
is the value of time for the commuter travelling in mode $i$,  $\nu_{C}$ is the average
speed in New Delhi under congested conditions, and $\nu_{f}$ is the free-flow
speed of traffic. Both $\Psi_i$ and $\mu_i$ are provided 
in Table \ref{table18}. Notice that, in New Delhi, the number of trips by buses is more 
than twice that for cars and two-wheelers. This is intuitive, since most cars and two-wheelers serve 
individual travel needs, and may be used just for commuting from home to work. However in the case of buses, they
are in use almost throughout the day because of the scheduled public transportation trips. Just by looking at 
this table, one can come to the conclusion that whatever legislation is to be passed, favoring buses might have an overwhelming positive effect. 

The average trip size, $\mu_i$ remains almost the same for all the three categories at around 10-11 km/trip. Fewer number of trips for cars and two-wheelers 
outlines the enormous potential that 
ride-sharing and similar initiatives can have, especially in the case of cars, where average occupancy is 
mostly less than 50\%. While potentials for improvement and reducing costs exist in all the three categories, it must be
observed that bringing about improvements in bus systems is considerably easier since most buses are state-owned. In the 
case of two-wheelers and cars, coordination among a large number of commuters may be essential, before being able to bring about considerable improvements.

In equation (\ref{eq:TCCPL}), $\Lambda_{i}$
represents the average occupancy for mode $i$. The average occupancy for buses is among the lowest in several similar works. For instance, \cite{singh} uses 
an occupancy rate as high as 85\%, which translates to roughly an average occupancy of 34 in a 40-seater bus. This is an important
point to note since taking higher average occupancies may considerably increase productivity loss costs.
The free-flow speed ($\nu_{f}$) is taken to be
40 km/h as in \cite{tyagi}, and $\nu_{C}$ is taken to be 22.2 km/h as in \cite{rites}. 

With (\ref{eq:TCCPL}) and Table \ref{table18}, the total
costs of congestion due to productivity losses are readily computed,
and are listed in Table \ref{table19}. Throughout this
study we have used the exchange rate of 1 US\$ = 60 INR (Indian Rupee). From the Table \ref{table20}, cars contribute more than 10\% of the total productivity
loss. As the number of cars is projected to grow rapidly, there could
be a severe detrimental effect not only on the car passengers, but
on all the other road users as well. Also note that the productivity losses for buses are the highest, because of its higher occupancy compared to other modes of transport. This is a good area for policy-makers to focus.

%
%
%
%
%
%

\begin{table}
\centering{}%
\begin{tabular}{|c||c|c|c|}
\hline 
Mode  & Trips per day  & Occupancy  & Trip Size (km)\tabularnewline
\hline 
\hline 
Car  & 2902120  & 2.2  & 11.28\tabularnewline
\hline 
Bus  & 7276892  & 20.0  & 10.66\tabularnewline
\hline 
Two-wheeler  & 3250755  & 1.2  & 10.03\tabularnewline
\hline 
\end{tabular}
\caption{Trips per day in New Delhi \cite{rites}}
\label{table18}
\end{table}

\begin{table}

\centering{}%
\begin{tabular}{|c||c|}
\hline 
Mode  & Cost (in million US\$/Yr)\tabularnewline
\hline 
\hline 
Car  & 869\tabularnewline
\hline 
Bus  & 6310\tabularnewline
\hline 
Two-wheeler  & 239\tabularnewline
\hline 
Total  & 7410\tabularnewline
\hline 
\end{tabular}

\caption{Total costs of congestion in New Delhi due to productivity losses}

\label{table19}
\end{table}

\subsection{Air Pollution Costs }

The computation of total congestion costs due to air pollution follows a 
comparison approach against the free-flow scenario. The factors considered to compute air pollution costs are similar to those introduced in the previous section. 
An important factor to be considered 
is the correction factor, that must provide an appropriate comparison with the baseline scenario. The correction 
factor must preferably be in terms of $\nu_C$ and $\nu_f$, since these are already available.

Keeping these in mind, the Total Cost of Congestion due to vehicular Emission of air
pollutants, TCCE is given by: 
\begin{equation}
\text{TCCE}=365\left(\frac{\nu_{f}}{\nu_{C}}-1\right)\sum_{i}\Biggl(\Psi_{i}\mu_{i}\Biggl(\sum_{k}\rho_{i}^{k}\delta^{k}\Biggr)\Biggr).\label{eq:TCCE}
\end{equation}

In the above equation, the inner summation with respect to the $k^{th}$ pollutant provides cost due to
pollutants (CO, HC, $\text{NO}_X$ and PM) emitted per vkm for the $i^{th}$ mode. This data is obtained 
from Tables \ref{table5}, \ref{table6}, \ref{table7}, \ref{table8} and \ref{table9}. Once this cost has been computed for the $i^{th}$ mode for 
all pollutants, the outer summation seeks to compute the total costs for all modes, throughout the year. Product of the cost
of emissions per vkm with the average total vkm traversed per day ($\Psi_i \mu_i$) will give cost of emissions
per day, for the $i^{th}$ mode. The outer summation over all modes, gives the Total Cost of Congestion due to Emissions (TCCE). 

Note that in expression (\ref{eq:TCCE}), we introduce a correction
factor, which was not present in (\ref{eq:TCCPL}). This factor accounts
for the reduced speed due to congestion. The basic assumption underlying the correction
factor is that the pollutants emitted increase proportionally with the increase in travel time. This is only 
an approximation since in most cases, the emission characteristics and constituents change as the vehicle 
speeds change. The changes are cumbersome and difficult to model. Keeping the assumption, the fractional 
change in time on road due to congestion is:

\begin{equation}
\text{Correction Factor} = \frac{\text{T}_C - \text{T}_f}{\text{T}_f},
\end{equation}
where $\text{T}_C$ is the time taken by a commuter to travel a given distance in congested 
conditions and $\text{T}_f$ is the time taken to travel the same distance in free-flow conditions. Since 
distances are the same in both congested and free-flow conditions, we have:

\begin{equation}
\text{T}_C\nu_C = \text{T}_f\nu_f. \label{simple_elas}
\end{equation}

Using the above and simplifying, we have:
\begin{equation}
\text{Correction Factor} = \frac{\nu_f}{\nu_C}-1.
\end{equation}

This is a rather simplified approach to computing the correction factor. More accuracy may be obtained by 
including travel demand elasticities \cite{graham}, since increased congestion increases travel impedance, reducing 
demand for travel. Then, the following must hold:

\begin{equation}
\text{T}_C\nu_C = \Pi_{TD}\text{T}_f\nu_f, \label{better_elas}
\end{equation}

where $\Pi_{TD}$ < 1 is a factor to account for the reduced travel demand. Due to data unavailability and complexity in computing elasticities, we will use \eqref{simple_elas} instead 
of the slightly more accurate \eqref{better_elas}. The correction factor is hence equal to the fractional increase in travel time due to congestion.
Using \eqref{simple_elas}, the total costs of congestion due to air
pollution is given in Table \ref{table20}.

\begin{table}
\centering{}%
\begin{tabular}{|c||c|}
\hline 
Mode  & Cost (in million US\$/Yr)\tabularnewline
\hline 
\hline 
Car  & 41\tabularnewline
\hline 
Bus  & 670\tabularnewline
\hline 
Two-wheeler  & 19\tabularnewline
\hline 
Total  & 730\tabularnewline
\hline 
\end{tabular}
\caption{Total costs of congestion in New Delhi due to emission of air pollutants}
\label{table20}
\end{table}

\subsection{Accidents}

As in the previous section, in which we compute the marginal costs due to accidents, we see in 
this section that accidents contribute a less significant component to total costs. However, in this
case, the total costs due to accidents includes an aggregate total cost incurred due to accidental events 
involving a range of seriousness levels. Computing the total costs of congestion is slightly more direct than computing the marginal costs of congestion due to accidents because the data available is already
in an aggregate form. In the previous section we found the cost per vkm only after finding the total costs. Then, the 
Total Cost of Congestion due to Accidents (TCCA) is given by:

\begin{equation}
\text{TCCA} = \left(\frac{\nu_{f}}{\nu_{C}}-1\right)\sum_i{{\sum_l{\epsilon_{il}}H_{l}}}.
\end{equation}

We also include the correction factor introduced in (\ref{eq:TCCE}). The form of the correction factor, follows the underlying 
assumption that increased road-time increases the probability of meeting with an accident. However, note that in this case the elasticities
of travel demand will not come into play. We assume, and with reason, that travel time does not have elastic dependencies - a commuter who can complete his travel sooner, will 
not stay on road, just to ensure that the entire time he expected to be on the road elapses. The treatment of elasticities
is far more complex in this case. Additionally, some studies have also found that commuters prefer constant 
travel time over varying travel times, where commuters may actually end up saving time on some days \cite{jong}.

The total cost due to accidents is provided in Table \ref{table21}.
The year-wise breakup of the number of accidents, as obtained from \cite{police},
is enumerated in Table \ref{table22}. Note that the number
of accidents is largely stable in the years 2008 through 2013. Accidents do not seem to follow 
any increasing or decreasing trend with observable parameters.
%
Also, we notice that fatal accidents contribute to most of the costs. Thus,
there is a compelling case to formulate and enforce very strict safety
norms, that reduce fatalities in road accidents.

\begin{table}
\begin{tabular}{|c||c|c|c|}
\hline 
Severity  & Events  & Cost per Accident (INR)  & Total (million US\$/Yr)\tabularnewline
\hline 
\hline 
Fatal  & 1778  & 1745600  & 41.48\tabularnewline
\hline 
Major  & 5619  & 311430  & 23.38\tabularnewline
\hline 
Minor  & 169  & 40916  & 0.09\tabularnewline
\hline 
Total  & 7566  & -  & 64.95\tabularnewline
\hline 
\end{tabular}
\centering{}\caption{Total costs due to accidents in New Delhi \cite{tcs,newbery}}
\label{table21}
\end{table}

\begin{table}
\centering{}%
\begin{tabular}{|c||c|}
\hline 
Year  & Number of Accidents\tabularnewline
\hline 
\hline 
2008  & 8435\tabularnewline
\hline 
2009  & 7516\tabularnewline
\hline 
2010  & 7260\tabularnewline
\hline 
2011  & 7280\tabularnewline
\hline 
2012  & 6937\tabularnewline
\hline 
2013  & 7566\tabularnewline
\hline 
\end{tabular}
\caption{Total number of accidents in New Delhi (New Delhi traffic police)}
\label{table22}
\end{table}

\subsection{Fuel Wastage}

Fuel wastage due to traffic delays leads to losses that can be traced
directly to traffic congestion. Some of the reasons for this fuel wastage due to congestion include:

\begin{itemize}

\item  Stalling at traffic signals.

\item Stalling and reduced speeds in traffic jams and diversions.

\item Reduced speeds at narrowing and tapering roads.

\item Reduced speeds at junctions, intersections and flyover extremes.

\end{itemize}

The basic approach to computing total fuel wastage costs includes computing reduced speeds at congested
intersections and finding the equivalent fuel consumption at these stalling points. The next step is
to assign the wasted fuel monetary costs. We do not compute the costs due to
fuel wastage in New Delhi as this has already been widely researched by several
governmental policy think-tanks. The earliest known estimate of fuel-wastage
in New Delhi was provided by the Central Road Research Institute, back
in 1996. The wastage was estimated at 300,000 US\$/day \cite{crri}. 

In \cite{parida}, a survey based approach is followed, by earmarking 12 intersections in New Delhi 
catering to varying traffic densities. The study estimates that the waste fuel 
cost is as high as 994.45 crores of Rupees per annum according to the 2008 price levels. Later in 2010, an independent study conducted by the Center for Transforming
India has pegged this cost at approximately 10 crores of Rupees per
day \cite{mess}. This figure will be used in our computations.

\subsection{Summary of Total Costs}
In this subsection, we present the total costs of congestion, having considered the various 
components that contribute to the total costs of congestion. Table \ref{table23} shows the total
cost of congestion in New Delhi per year, with most of the data used to compute
the contributions of the underlying factors falling in the range of
2008-2010. In INR terms, traffic congestion costed New Delhi close to 54,000 crores of Rupees in the year 2013. There are a few important points that are to be emphasised as evident from Table \ref{table23}:

\begin{itemize}

\item Buses are the largest contributors to the total costs of congestion. But, considering the number
of trips per day that buses in New Delhi undertake and the number of commuters whose travel demands they 
satisfy, buses are probably the most efficient transportation means, in terms of total costs.

\item Costs due to productivity losses are the largest contributor to the total 
costs of congestion. All other factors fall within 10\% of the contribution made 
by costs due to productivity losses. 

\item The contributions by cars to congestion costs is almost a
billion US\$/yr and given the occupancy of less than 50\% these costs are 
likely to have the most potential for reduction.
\end{itemize}

\begin{table}
\centering{}%
\begin{tabular}{|c||c|}
\hline 
Mode  & Total Cost (in million US\$/yr)\tabularnewline
\hline 
\hline 
Car  & 911\tabularnewline
\hline 
Bus  & 6980\tabularnewline
\hline 
Two-wheeler  & 258\tabularnewline
\hline 
Accidents  & 64\tabularnewline
\hline 
Fuel Wastage  & 699\tabularnewline
\hline 
Total  & 8912\tabularnewline
\hline 
\end{tabular}
\caption{Total costs of congestion in New Delhi}
\label{table23}
\end{table}

\section{Cost Projections for Productivity Losses\protect \protect \protect
\protect \\
 and Air Pollution}

In Section II, we computed the marginal costs of congestion, followed by total costs of congestion in Section III. An important
requirement now is to be able to approximately tell how these costs are expected to change with time. This is
an essential requirement since it justifies recommended  infrastructural spending to ease congestion.

This section provides the cost projections for the marginal
and total costs until 2030. The closest work relating to the results in this section 
is \cite{kokaz}, which uses projections to determine the optimal transportation mix. In our case, we use the projected vehicle population growth to determine both marginal and total costs of 
congestion and in turn, make projections on these costs. 
%
%
%
%


Of the four underlying factors that have been considered as contributors to total costs of congestion, we argue
that two of the factors - fuel wastage costs and accident costs may be neglected. Projections will then be made for the productivity losses and air pollution costs. 

Accident costs are neglected in making the projections because:
\begin{itemize}

\item The number of accidents are difficult to be modelled and predicted. The number of accidents does not seem 
to show any strong dependence on the number of vehicles \cite{police}.
\item Even if a method to accurately determine number of accidents was perfected, the contribution from such 
costs would be dwarfed by the total costs of congestion. For instance, in the present scenario, the contribution to the
total costs from accidents is just about 0.72\% of the total. A similar situation is encountered in the case of marginal costs.

\end{itemize}

The fuel wastage costs are also neglected for the following reasons:
\begin{itemize}
\item The magnitude of fuel wastage costs is distorted due to changes in global oil prices. Projecting fuel wastage 
requires being able to project oil prices many years hence, which is an impossible task \cite{regnier}.

\item The dependence of fuel quantity wasted with the number of vehicles is non-trivial and may strongly depend on 
several factors such as infrastructure and other network characteristics of the transportation network.

\item This is another minor contributor to the total costs, presently contributing less than 8\% of the total costs and can hence be neglected. 

\end{itemize}

Projections of marginal and total costs of congestion are made by obtaining the growth projections of the two underlying factors affecting these
costs - productivity losses and air pollution costs. However, making projections directly based on these two factors is non-trivial. A good approach would be 
to find a common dependence on which both these two factors depend, and for which plenty of past data is available so as to make the statistical projections meaningful. Vehicular population is one such common dependence and it satisfies the past data availability criteria also.


There are advantages in making projections for productivity losses
and vehicular emissions indirectly based on the projections for vehicle
population, rather than directly making projections based on the individual
factors:
\begin{itemize}

\item In the indirect approach, the projections are independent
of the model used to arrive at the contributions made by productivity
loss and vehicular emissions based costs. 

\item Another inherent advantage
is that projections on the vehicular population of New Delhi have been
widely studied; however, this is not true of the individual factors.
\end{itemize}

Note that this approach lacks accuracy, as with increasing vehicular populations, network parameters 
describing the network characteristics may well change. Though the approach underestimates the projected costs, it will
serve to justify minimal infrastructural spending. The projections for vehicular population in New Delhi is completed using a 
spreadsheet model. This completes the first step of the projection process.
 
The next task is to model the dependence between vehicular population and the two most relevant underlying
factors making up marginal and total costs. The equations below capture the dependence of these two factors on
vehicle population. From (\ref{eq:speed-flow}), the dependence of
Productivity Loss Costs (PLC) on the vehicle population, $N$, is: 
\begin{equation}
\text{PLC}\varpropto e^{A_{4}N}.\label{eq:PLC}
\end{equation}
Similarly, since the Vehicular Emission Costs (VEC) depend on the
number of vehicles, assuming an equal distribution and a similar modal
share throughout the projected years, we have: 
\begin{equation}
\text{VEC}\varpropto N.\label{eq:VEC}
\end{equation}

\begin{table}
\centering{}%
\begin{tabular}{|c||c|c|c|}
\hline 
Year  & Two-wheeler  & Car  & Bus \tabularnewline
\hline 
\hline 
2015  & 4918777  & 2512234  & 64748 \tabularnewline
\hline 
2018  & 5608980  & 2885110  & 74713 \tabularnewline
\hline 
2020  & 6033646  & 3127639  & 84643 \tabularnewline
\hline 
2023  & 6634911  & 3461118  & 89259 \tabularnewline
\hline 
2025  & 7013511  & 3693622  & 93971 \tabularnewline
\hline 
2027  & 7402890  & 3908506  & 99580 \tabularnewline
\hline 
2030  & 8056069  & 4236245  & 109330 \tabularnewline
\hline 
\end{tabular}
\caption{Vehicular population projections in New Delhi}
\label{table24}
\end{table}

The projections for vehicular population in New Delhi obtained from the simple spreadsheet model are 
provided in Table \ref{table24}. Using Table \ref{table24} and equations (\ref{eq:PLC},\ref{eq:VEC}),
the projections for the marginal costs of congestion are as in Table
\ref{table25}. Similarly, projections for the total costs
of congestion are given in Table \ref{table26}.

An assumption regarding the projection for buses is that the government will continue 
sanctioning buses in line with the demands of the population, and will not look to increase bus frequencies so as to lower average occupancies. This 
is only a slight underestimation, since with increasing living standards in India, it is highly likely that buses will be sanctioned at an higher rate 
than predicted. Though this effect will not affect the productivity loss costs (which depends only on the number of passengers), it will
increase the environmental costs due to the lower average occupancies, and hence increased vkm per day. This will once again underestimate the cost projections.

An alarming observation based on the projections is the nearly 70\% increase in the number
of cars. Clearly, this will not be sustainable, and will have a damaging
impact, particularly on productivity losses, environmental costs,
and fuel wastage.

\begin{table}
\centering{}%
\begin{tabular}{|c||c|c|c|}
\hline 
Year  & Car  & Bus  & Two-wheeler \tabularnewline
\hline 
\hline 
2015  & 11.16  & 21.69  & 2.61\tabularnewline
\hline 
2018  & 13.93  & 23.20  & 3.89\tabularnewline
\hline 
2020  & 16.08  & 24.79  & 4.99\tabularnewline
\hline 
2023  & 19.61  & 25.55  & 7.11\tabularnewline
\hline 
2025  & 22.51  & 26.35  & 8.88\tabularnewline
\hline 
2027  & 25.57  & 27.32  & 11.18\tabularnewline
\hline 
2030  & 31.07  & 29.08  & 16.47\tabularnewline
\hline 
\end{tabular}
\caption{Cost projections - marginal costs of congestion (INR/vkm)}
\label{table25}
\end{table}

\begin{table}

\centering{}%
\begin{tabular}{|c||c|c|c|c|}
\hline 
Year  & Car  & Bus  & Two-wheeler  & Total\tabularnewline
\hline 
\hline 
2015  & 1033  & 7233  & 331  & 8597\tabularnewline
\hline 
2018  & 1288  & 7746  & 493  & 9527\tabularnewline
\hline 
2020  & 1486  & 8282  & 630  & 10398\tabularnewline
\hline 
2023  & 1809  & 8540  & 896  & 11245\tabularnewline
\hline 
2025  & 2074  & 8809  & 1120  & 12003\tabularnewline
\hline 
2027  & 2354  & 9138  & 1410  & 12902\tabularnewline
\hline 
2030  & 2857  & 9731  & 2070  & 14658\tabularnewline
\hline 
\end{tabular}
\caption{Cost projections - total costs of congestion (million US\$/yr)}
\label{table26}
\end{table}

\section{Conclusions and Recommendations}

The key takeaways from our study are summarised below. 
\begin{itemize}
\item After monitoring the taxi GPS traces on the roads of New Delhi for a period of over an year, we noticed that there is a negative trend in the average taxi speed. We also observed a positive trend in the number of taxis during the same period of study. These patterns point towards the increasing levels of congestion in the city.

\item The results from the K-S test suggest that the speed distributions for the years 2013 and 2014 are dissimilar, and that the taxi speeds are statistically higher in 2013. The reduction in speed in 2014 may lead to more productivity losses, pollution losses and fuel wastages, compared to 2013. Hence, it is very likely that the total congestion costs in 2014, and the subsequent years, will be higher than that in 2013. This supports the cost projections in Table \ref{table26}. 
 
\item Even 15 years after the authors' claim in \cite{kokaz}, buses still are the most popular means of road transport catering to about
60\% of New Delhi's total demand. The state-owned New Delhi Transport Corporation
buses are in fact the largest CNG-driven fleet in the world \cite{plan}.
It is clear from our study that buses are contributing a substantial
portion of the total costs, primarily due to productivity losses.
The productivity loss due to congestion delays of commuters who use
buses accounts for about 75\% of total costs of congestion. 

\item Idling at traffic lights, signalised intersections and busy junctions  due to congestion causes fuel wastage, which is another
source of substantial costs. With the number of cars projected to
increase sharply, this component is expected to play an increasingly
significant role. 

\item From Table \ref{table25}, we see that the projected marginal
congestion costs of cars approach that of buses. This means that in
the year 2030, according to our projections, the cost of adding a
vkm of car travel to the existing traffic network is very similar
to the cost of adding a vkm of bus travel to the same network, despite
the enormous differences in size and hence in road space occupancy.
This goes to show that the New Delhi traffic network will be so saturated
that the addition of one vkm of bus or car will be viewed similarly. 

\item Another important conclusion is that cars have the most potential for cost savings, due to two important
reasons: average occupancies not exceeding 50\% and a low number of average trips per day. Ride-sharing 
and similar arrangements in New Delhi will have tremendous potential in terms of cost saving as well as easing congestion.

\item The economic costs arising from accidents is not a significant proportion
of the total costs. Though accidents entail significant and irrevocable personal losses, their contribution
is less significant
from a macroeconomic perspective. 
\end{itemize}
Based on the results obtained so far and the conclusions above, we provide some key recommendations to address the issues
identified. 
\begin{itemize}
\item The Government should look into setting up dedicated bus lanes. This
would considerably reduce the productivity losses for commuters who
use buses, encouraging other private transport users to commute by
buses due to the reduced transit time. Our study also adds strong
credibility to various works in literature that make a case for dedicated
lanes for buses in New Delhi \cite{jain,sarma}. In order to make dedicated
bus lanes effective, it would be important to have more frequent,
and more comfortable buses. This could also help in shifting a fraction
of the motorists to buses. 

\item Employ
state-of-the-art scheduling policies for buses. There is also
a case to be made for equipping public buses with GPS and making
the data publicly available. This would enable real-time solutions
and innovation to flourish. Though these recommendations entail additional
spending on the part of the Government, public transport investments
by the Government in New Delhi have had high returns, as is evident in
the case of the New Delhi metro \cite{murty}.

\item As fuel wastage is expected to increase, it would be important to
employ intelligent traffic management systems, including smart traffic
lights. Such solutions could be extremely valuable in future smart
cities, where it may be possible to install the required infrastructure
in advance. 

\item Car pooling and other similar measures must be promoted, and the Government
should help facilitate and incentivise such practices where ever possible. 

\end{itemize}
\begin{table}
\centering{}%
\begin{tabular}{|l||l|}
\hline 
\textit{Policy Recommendation}  & \textit{Cost Impact} \tabularnewline
\hline 
\hline 
Dedicated Bus Lanes  & 6300 million US\$/yr\tabularnewline
\hline 
Strict Vehicular Emission Control Norms  & 730 million US\$/yr\tabularnewline
\hline 
Safety and Accident Prevention Features  & 65 million US\$/yr\tabularnewline
\hline 
\end{tabular}
\caption{Policy recommendations and likely impact cost}
\label{table27}
\end{table}

With regards to future work, a more comprehensive study on all the aspects
that impact costs of congestion in New Delhi is certainly required. There are several aspects of this study, especially
in the computational modelling aspects that can be extended. Some of these among several others are:

\begin{itemize}

\item Include travel demand elasticities to obtain a more accurate correction factor in \eqref{simple_elas}.

\item Compute the new $\delta^k$ costs for New Delhi. The existing costs are fairly outdated, last computed for the year
1998 using the transfer of benefit method as used in \cite{mandal}.

\item Projections can be made more accurately, by considering parametric changes that are influenced by the vehicular population. The present
approach makes projections based on vehicular growth projections, but assumes all else to be constant, hence underestimating the cost projections.

\item Recompute the fitting parameters $A_{1j}, A_2, A_3 \text{ and } A_4$ obtained from \cite{sen}. The parameters are expected to have slightly changed now, due to
the passage of time since they were first computed using curve-fitting methods in 2010.

\end{itemize}

The advantages of replicating similar systematic study in other
major Indian cities can be clearly seen. Such studies would better inform cost-benefit
considerations for the numerous possible solutions that may be considered,
towards providing a smarter transportation infrastructure in various
cities, which is an important requirement in the developing world.

\section*{Acknowledgements}

The work is undertaken as part of an ITRA, Media Lab Asia, project
entitled ``De-congesting India's transportation networks.'' The authors are also thankful to Mr. Gopal Krishna Kamath for the helpful discussions.

\end{document}